\documentclass{emulateapj}

\shorttitle{Extremely broad RRL masers toward HW2}
\shortauthors{I. Jim\'enez-Serra et al.}

\begin{document}

\title{Extremely broad radio recombination maser lines toward the high-velocity 
ionized jet in Cepheus A HW2}

\author{I. Jim\'{e}nez-Serra\altaffilmark{1}, J. Mart\'{\i}n-Pintado\altaffilmark{2}, 
A. B\'aez-Rubio\altaffilmark{2}, N. Patel\altaffilmark{1} and C. Thum\altaffilmark{3}}

\altaffiltext{1}{Harvard-Smithsonian Center for Astrophysics, 
60 Garden St., Cambridge, MA 02138, USA; ijimenez-serra@cfa.harvard.edu; npatel@cfa.harvard.edu}

\altaffiltext{2}{Centro de Astrobiolog\'{\i}a (CSIC/INTA),
Ctra. de Torrej\'on a Ajalvir km 4,
E-28850 Torrej\'on de Ardoz, Madrid, Spain; 
jmartin@cab.inta-csic.es,baezra@cab.inta-csic.es}

\altaffiltext{3}{Institut de Radio Astronomie Millim\'etrique, 300 Rue 
de la Piscine, F-38406 St. Martin d'H\`eres, France; thum@iram.fr}

\begin{abstract}

We present the first detection of the H40$\alpha$, H34$\alpha$ and 
H31$\alpha$ radio recombination lines (RRLs) at millimeter wavelengths 
toward the high-velocity, ionized jet in the Cepheus A HW2 star forming region. From our
single-dish and interferometric observations, we find that the measured RRLs 
show extremely broad asymmetric line profiles with zero-intensity linewidths of 
$\sim$1100$\,$km$\,$s$^{-1}$. From the linewidths, we estimate a terminal
velocity for the ionized gas in the jet of $\geq$500$\,$km$\,$s$^{-1}$, consistent with that obtained from the proper motions of the HW2 radio jet. The total integrated line-to-continuum flux ratios of the H40$\alpha$, H34$\alpha$ and H31$\alpha$ lines are 43, 229 and 280$\,$km$\,$s$^{-1}$, 
clearly deviating from LTE predictions. These ratios are very similar to those observed for the RRL maser toward MWC349A, suggesting 
that the intensities of the RRLs toward HW2 are affected by maser emission. Our radiative transfer modeling of the RRLs shows that their asymmetric profiles could be explained by maser emission arising from a bi-conical
radio jet with a semi-aperture angle of 18$^\circ$, electron density
distribution varying as r$^{-2.11}$ and turbulent and expanding wind velocities
of 60 and 500$\,$km$\,$s$^{-1}$.

\end{abstract}


\section{Introduction}

Radio recombination lines (RRLs) are excellent probes of the kinematics of the 
ionized gas in ultracompact (UC) HII regions \citep[][]{gar89,chu89}.
These lines typically show simple Gaussian profiles with linewidths of 
25-30$\,$km$\,$s$^{-1}$, attributed to unresolved gas motions 
and/or pressure broadening \citep[][]{gau95,aff96,ket08}.

In addition to the {\it classical} UC HII regions with simple narrow 
($\sim$30$\,$km$\,$s$^{-1}$) Gaussian RRLs, \citet{jaf99} reported 
that 30\% of the observed UC HII regions show even broader RRL emission with linewidths of 70-200$\,$km$\,$s$^{-1}$. 
These sources have power law continuum spectra with spectral indeces
$\sim$0.6 \citep[characteristic of constant velocity stellar winds;][]{oln75},
and elongated/bipolar morphologies resembling ionized flows. 
\citet{jaf99} proposed that broad RRL emission could arise from    
bipolar ionized winds generated in photo-evaporating neutral disks \citep{hol94,gor09}.

MWC349A is the best studied object with broad H recombination lines at optical, IR, centimeter and millimeter 
wavelengths \citep{har80,ham86,alt81,mar89}. This source, which has a rotating edge-on 
disk with a bipolar ionized flow \citep[][]{coh85},
is unique in its category because 
its RRLs at $\lambda$$\leq$2$\,$mm (quantum numbers n$<$35) are masers 
\citep[optical depths $<$-1;][]{mar89,mar94}. 
The RRL maser spots are located on the ionized surface of the 
disk \citep{pla92,wei08,mar11}, where the densities of the ionized 
gas are high enough ($\geq$10$^6$$\,$cm$^{-3}$) to invert the population 
of the levels involved in the RRL \citep[][]{wal90}. Although
these masers are expected to be found in other UC HII regions \citep[][]{mar02}, 
MWC349 is the only object where this emission has been reported so far  
in star forming regions. 

We present the first detection of RRL maser emission toward the 
Cepheus A HW2 high-mass star forming region. This source 
shows a collimated, high-velocity ionized jet, with a continuum 
spectral index $\sim$0.6 \citep[][]{rod94}. Like MWC349A, the 
molecular material around HW2 is mainly distributed in a neutral circumstellar disk \citep[][]{pat05,jim09} that seems to be 
photo-evaporating \citep{jim07}. The proper motions of the HW2 radio jet  
suggest an expanding velocity for the outflowing gas of 
$\geq$500$\,$km$\,$s$^{-1}$ \citep{cur06}. This is consistent with 
the extremely broad line profiles (zero-intensity linewidths of 
$\sim$1100$\,$km$\,$s$^{-1}$) of the H40$\alpha$, H34$\alpha$ and 
H31$\alpha$ maser lines detected in our single-dish and 
interferometric observations. 
Cepheus A HW2 is the second RRL maser object detected 
to date in star forming regions.   

\section{Observations}
\label{obs}

\begin{figure*}
\begin{center}
\includegraphics[angle=270,scale=.8]{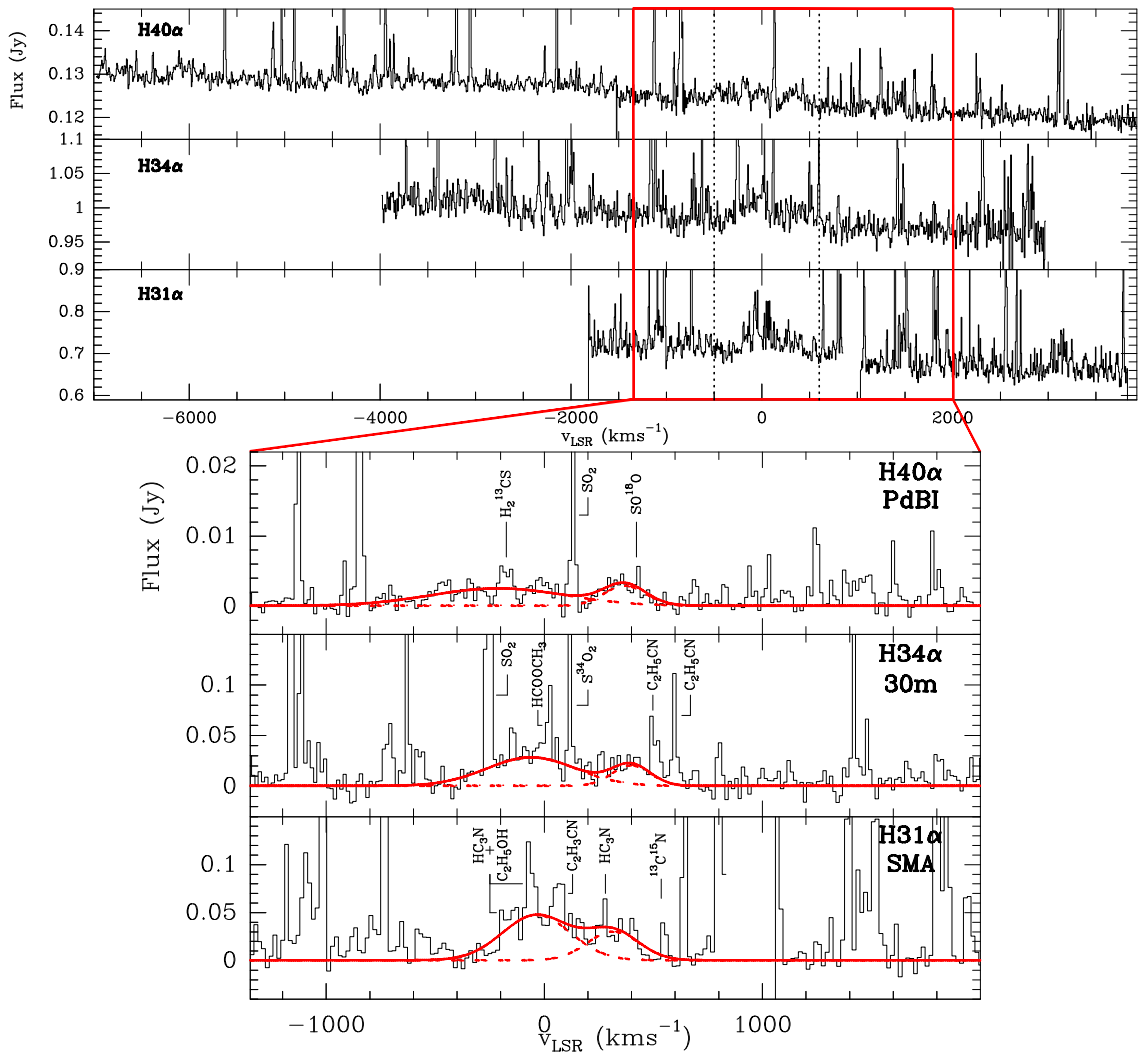}
\caption{{\it Upper panels:} Full spectra of the 
H40$\alpha$, H34$\alpha$, and H31$\alpha$ RRLs measured 
toward Cepheus A HW2 with the PdBI, 30$\,$m and SMA at 
3, 1.9 and 1.4$\,$mm, respectively. Box shows the
region of the RRL spectra shown in the 
lower panels of this Figure. Dotted lines indicate the 
extension in velocity from $v_{LSR}$$\sim$$-$500 to 
600$\,$km$\,$s$^{-1}$ of the H40$\alpha$, H34$\alpha$, 
and H31$\alpha$ RRLs. 
{\it Lower panels:} Zoom-in of the continuum-subtracted
RRLs spectra, after being smoothed
to a velocity resolution of $\sim$15-17$\,$km$\,$s$^{-1}$. 
Identifications of some molecular lines
in the RRL spectra are shown. We also report 
the individual Gaussian fits to the blue-shifted and red-shifted 
velocity components of the RRL emission (dashed lines), 
and the total Gaussian fit to these lines (solid lines; 
see also Table$\,$\ref{tab1}).\label{f1}}
\end{center}
\end{figure*}

The H40$\alpha$ line at $\sim$99$\,$GHz was observed toward 
Cepheus A HW2 with the IRAM Plateau de Bure Interferometer 
(PdBI) in the D configuration (beam of 
5.0$''$$\times$4.5$''$, P.A.=91$^\circ$)\footnote{Based on 
observations carried out with the IRAM 30$\,$m and 
Plateau de Bure Interferometer. IRAM is supported by 
INSU/CNRS (France), MPG 
(Germany) and IGN (Spain)}. The phase center was 
set at $\alpha(J2000)$=22$^{h}$56$^{m}$17.98$^s$, 
$\delta(J2000)$=+62$^{\circ}$01$'$49.5$''$. The wide-band 
correlator WideX provided a continuous frequency coverage of 
3.6$\,$GHz, and a spectral resolution of 1.95$\,$MHz (i.e. 
a velocity coverage and resolution of $\sim$10900$\,$km$\,$s$^{-1}$ 
and $\sim$6$\,$km$\,$s$^{-1}$, respectively). 
3C454.3 (23$\,$Jy) and 3C273 (12$\,$Jy) were used as
bandpass calibrators.
We observed MWC349 (1.2$\,$Jy) and 1749+096 (2$\,$Jy) as flux 
calibrators, and 0016+731 (0.6$\,$Jy) and 2146+608 
(0.3$\,$Jy), as phase calibrators. Data reduction, calibration, 
imaging and cleaning were performed with 
the GILDAS software\footnote{See http://www.iram.fr/IRAMFR/GILDAS.}.  

The H34$\alpha$ line at $\sim$160$\,$GHz,
was detected toward HW2 with the IRAM 30$\,$m telescope in a single-pointing 
observation during 140$\,$min. The EMIR E1 receiver was tuned to 
single sideband (SSB) with rejection of $\geq$10$\,$dB. The beam
size was 15$''$. The wide-band auto-correlator WILMA provided 
a total bandwidth and spectral resolution of 4$\,$GHz and 2$\,$MHz 
($\sim$6700$\,$km$\,$s$^{-1}$ and 3.7$\,$km$\,$s$^{-1}$, respectively). 
Typical system temperatures were 200-225$\,$K. Intensities were 
calibrated in $T_A^*$, and converted into total flux (Jy) by using 
$S/T_A^*$=6.4$\,$Jy/K 
\footnote{See http://www.iram.es/IRAMES/mainWiki/Iram30mEfficiencies.}. 

The H31$\alpha$ line at 210$\,$GHz was imaged toward HW2 with 
the Submillimeter Array (SMA) in the subcompact configuration 
(beam of 4.8$''$$\times$3.8$''$, P.A.=21$^\circ$)\footnote{The 
Submillimeter Array is a joint project between the Smithsonian 
Astrophysical Observatory and the Academia Sinica Institute of 
Astronomy and Astrophysics and is funded by the Smithsonian 
Institution and the Academia Sinica.}. The correlator setup 
provided a total bandwidth of 4$\,$GHz per side-band, 
and a spectral resolution of 0.8$\,$MHz. This corresponds 
to a velocity coverage of $\sim$5700$\,$km$\,$s$^{-1}$ and a 
velocity resolution of 1.1$\,$km$\,$s$^{-1}$. 
3c279 was used as bandpass calibrator; MWC349, as flux calibrator 
(1.8$\,$Jy); and 0102+584 (1.1$\,$Jy) and BLLAC (5$\,$Jy), as gain calibrators. Data calibration was performed within the MIR IDL package, 
while imaging and cleaning was done with MIRIAD.

\section{Results}
\label{res}

\begin{deluxetable*}{lccccccccc}
\tabletypesize{\scriptsize}
\tablecaption{Derived parameters of the H40$\alpha$, H34$\alpha$ and H31$\alpha$ 
RRLs toward Cepheus A HW2.\label{tab1}}
\tablewidth{0pt}
\tablehead{
\colhead{RRL} & \colhead{$\nu$ (MHz)} & \colhead{$\lambda$ (mm)} & & \colhead{$T_L$ (Jy)} & \colhead{$v_{LSR}$ (km$\,$s$^{-1}$)} &
\colhead{$\Delta v$ (km$\,$s$^{-1}$)} & \colhead{$T_L$$\Delta v$ (Jy$\,$km$\,$s$^{-1}$)} & \colhead{$S_{\nu}^{tot}$ (Jy)\tablenotemark{a}} & \colhead{$S_{\nu}^{ff}$ (Jy)\tablenotemark{b}}
}
\startdata

H40$\alpha$ & 99022.96 & 3.0 & B & 0.0029 (0.0010)\tablenotemark{c} & $\sim$$-$210 & $\sim$620 & 1.9 (0.1)\tablenotemark{d} & 0.12 & 0.059 \\
& & & R & 0.0034 (0.0010) & $\sim$368 & $\sim$190 & 0.68 (0.05) & & \\ \hline

H34$\alpha$ & 160211.52 & 1.9 & B & 0.033 (0.009) & $\sim$$-$60 & $\sim$410 & 14.4 (0.7) & 0.37\tablenotemark{e} & 0.082 \\
& & & R & 0.024 (0.009) & $\sim$400 & $\sim$180 & 4.4 (0.5) & & \\ \hline

H31$\alpha$ & 210501.77 & 1.4 & B & 0.055 (0.006) & $\sim$$-$36 & $\sim$313 & 18.5 (0.4) & 0.69 & 0.099 \\
& & & R & 0.035 (0.006) & $\sim$316 & $\sim$240 & 8.9 (0.4)

\enddata

\tablenotetext{a}{Total continuum (dust + free-free) emission flux measured toward HW2.}
\tablenotetext{b}{Free-free continuum flux obtained by assuming a spectral index of $\sim$0.6, as derived from the VLA data of \citet{rod94}.}
\tablenotetext{c}{The error in the RRL peak intensity corresponds to the 1$\sigma$ noise level in the spectra (1$\,$mJy, 9$\,$mJy and 6$\,$mJy 
for the H40$\alpha$, H34$\alpha$ and H31$\alpha$ RRLs respectively; 
see lower panels in Figure$\,$\ref{f1}).}
\tablenotetext{d}{The error in the integrated intensity flux of the RRLs is calculated as $\sigma_{Area}$=1$\sigma$$\times$$\sqrt{\delta v\times\Delta v}$, with $\delta v$ the velocity resolution of
$\sim$15-17$\,$km$\,$s$^{-1}$ in the RRL spectra (see Section$\,$\ref{res}), and $\Delta v$ the linewidth derived in the Gaussian fit
(Column 7 in this Table).}
\tablenotetext{e}{Derived from the continuum flux measured with 
the PdBI and SMA at 3$\,$mm and 1.4$\,$mm .}

\end{deluxetable*}

In Figure$\,$\ref{f1} (upper panels), we present the 
full spectra measured toward Cepheus A HW2 at the wavelengths 
of the H40$\alpha$, H34$\alpha$ and H31$\alpha$ RRLs, and smoothed to 
a velocity resolution of $\sim$6$\,$km$\,$s$^{-1}$. Since the 
HW2 radio jet is unresolved in the PdBI and SMA images 
\citep[size of $\leq$2$''$;][]{rod94,cur06}, the H40$\alpha$ 
and H31$\alpha$ spectra were averaged within the 5$"$-beam 
of the PdBI and SMA observations. The measured RRL spectra 
(Figure$\,$\ref{f1}) show three 
different features: i) a strong slope due to the increase of the dust continuum emission with frequency ($S_{\nu}\propto\nu^\alpha$); ii) a {\it forest} of narrow molecular lines arising mainly from the HC source in HW2 \citep[][]{mar05,jim09}; 
and iii) three faint and extremely broad features extending 
in velocity from $\sim$-500$\,$km$\,$s$^{-1}$ to 
$\sim$600$\,$km$\,$s$^{-1}$ (zero-intensity linewidths 
of $\sim$1100$\,$km$\,$s$^{-1}$). 

From the slope in the RRL spectra, we 
derive spectral indeces for the continuum emission 
of $\sim$2, consistent with that calculated by 
\citet[][]{com07}. The PdBI and SMA indeces were obtained 
after subtracting the SEDs of the bandpass calibrators 
3C454.3, 3C273 and 3C279. 
The derived continuum (dust+free-free) level at the frequencies 
of the H40$\alpha$ and 
H31$\alpha$ RRLs is $\sim$0.12 and 0.69$\,$Jy, respectively. 
For the H34$\alpha$ line, this level was affected 
by an anomalous diffraction within the E1 receiver, although 
this did not affect the observed continuum slope and molecular 
line flux. The 2$\,$mm continuum level 
($\sim$0.37$\,$Jy) is estimated from the 
measured 3$\,$mm and 1.4$\,$mm continuum fluxes (Column$\,$9 in 
Table$\,$\ref{tab1}).

From Figure$\,$\ref{f1}, it is clear that molecular line 
confusion becomes an issue in this kind of studies, since multiple
line blending could generate broad features mimicing 
broad emission from RRLs. Molecular lines from 
species such as SO$_2$, C$_2$H$_5$CN, 
HC$_3$N or C$_2$H$_5$OH are indentified at frequencies close 
to the RRLs (lower panels in Figure$\,$\ref{f1}). However, their linewidths are narrow ($\sim$4-8$\,$km$\,$s$^{-1}$), and the 
resulting blending features do not exceed 
$\sim$100$\,$km$\,$s$^{-1}$. Note that other blending features are also 
found across the RRLs spectral bands, with similar widths. 
It is then unlikely that line blending is resposible for the 
extremely broad emission, with zero-intensity linewidths 
of $\sim$1100$\,$km$\,$s$^{-1}$, detected toward Cepheus A HW2. 
The probability to detect three similar (and extremely broad) 
features as a result of line confusion at three different frequency ranges, is also very small. We thus conclude that 
these features are likely associated with the emission of the 
H40$\alpha$, H34$\alpha$ and H31$\alpha$ RRLs, formed in the 
high-velocity HW2 ionized jet (Section$\,$\ref{mod}).  

Figure$\,$\ref{f1} (lower panels) shows a zoom-in of the continuum-subtracted RRL spectra, 
smoothed to a velocity resolution of 15-17$\,$km$\,$s$^{-1}$. The continuum
emission was subtracted by fitting a polynomial function of order 1. 

Although less clear for the H40$\alpha$ line, 
the extremely broad RRL profiles are asymmetric 
and have a similar kinematical structure with two broad (red- and 
blue-shifted) velocity components. The two-component Gaussian
fits of this emission are shown in Figure$\,$\ref{f1} (see solid and 
dashed lines). The derived parameters for the red- (R) and 
blue-shifted (B) components are given in 
Table$\,$\ref{tab1}. 

Table$\,$\ref{tab1} also reports 
the contribution from only the free-free continuum emission at
3, 1.9 and 1.4$\,$mm (Column$\,$10), derived by 
extrapolating the HW2 fluxes at centimeter wavelengths and by assuming an spectral index $\sim$0.6 \citep[][]{rod94}. These values, however,   
should be considered as upper limits since the actual free-free 
emission at millimeter wavelengths could be smaller than reported in
Table$\,$\ref{tab1} \citep[][]{com07}. If overestimated, these fluxes 
would imply an even stronger maser amplification effect 
for the observed RRLs (Section$\,$\ref{ltr}).

The derived integrated intensities of the RRLs lie above the 
9$\sigma_{Area}$ level (Column$\,$8), with $\sigma_{Area}$ 
the integrated intensity error in the spectra (see caption in 
Table$\,$\ref{tab1}). The derived peak and integrated intensities not only 
show a systematic trend to increase for increasing frequency (or
decreasing quantum number n) for both velocity components, but 
to be brighter for the blue-shifted emission compared to the 
red-shifted component. This asymmetry in the RRL profiles can be 
explained as RRL maser emission in the HW2 jet (Section$\,$\ref{mod}).  


Ignoring the results from the H40$\alpha$ blue-shifted 
component (its peak velocity and linewidth are factors of 
$\sim$4 and 2 larger than for H34$\alpha$ and H31$\alpha$; 
Table$\,$\ref{tab1}), the averaged peak velocities and 
linewidths are, respectively, $-$50$\,$km$\,$s$^{-1}$ and 
360$\,$km$\,$s$^{-1}$ for the blue-shifted emission, 
and 360$\,$km$\,$s$^{-1}$ and 
200$\,$km$\,$s$^{-1}$ for the red-shifted gas. 

\section{Line-to-Continuum flux ratios: Radio Recombination Line Masers in HW2}
\label{ltr}

\citet{mar89} showed that the generation of RRL emission at millimeter 
wavelengths can suffer from non-LTE effects that lead to the formation of RRL 
masers in ionized winds. Since Cepheus A HW2 shows similar properties 
to the RRL maser object MWC349A, it is possible that the RRLs toward HW2 also form under non-LTE conditions.  

In Section$\,$\ref{res}, we have shown that there is a trend for the 
RRL peak and integrated intensities to increase with frequency. This 
behavior is expected for RRL emission, since the integrated 
line-to-continuum flux ratios for RRLs \citep[ILTRs, defined as 
$\Delta v$$T_L$$/$$T_C$;][]{mar89} increase with frequency as 
$\nu^{1.1}$ for optically thin emission and LTE conditions. From Table$\,$\ref{tab2} 
(Columns$\,$2 and 3), however, we find that the measured ILTRs for 
these lines clearly deviate from the LTE predictions. 
The RRLs detected toward HW2 are therefore affected by maser effects. 
Like MWC349A (Column$\,$4 in Table$\,$\ref{tab2}), the deviations from 
LTE for the ILTR of the H40$\alpha$ line are less prominent than 
those of H34$\alpha$ and H31$\alpha$ at higher frequencies. This is 
explained by the fact that the $\beta_n$ factors \citep[the 
$\beta_n$ factor is proportional to the effective absorption 
coefficient of the RRL;][]{wal90} are expected to be more negative 
for RRLs with n$\leq$35 than for those with n=40, 
at the electron densities of $\sim$1-5$\times$10$^4$$\,$cm$^{-3}$ 
derived toward HW2 \citep[][]{cur06}. 

\begin{deluxetable}{lccc}
\tablecaption{Integrated Line-to-Continuum flux ratios (ILTRs).\label{tab2}}
\tablewidth{0pt}
\tablehead{
\colhead{RRL} & \colhead{HW2\tablenotemark{a} (km$\,$s$^{-1}$)} & \colhead{LTE\tablenotemark{b} (km$\,$s$^{-1}$)} & \colhead{MWC349 (km$\,$s$^{-1}$)}
}
\startdata

H40$\alpha$ & 43 & $\sim$30 & 39\tablenotemark{c} \\ 
H34$\alpha$ & 229 & $\sim$43 & 110\tablenotemark{d} \\ 
H31$\alpha$ & 280 & $\sim$58 & 215\tablenotemark{c}

\enddata

\tablenotetext{a}{Derived from the results shown in Columns$\,$8 and 10 of Table$\,$\ref{tab1}.}
\tablenotetext{b}{Calculated for optically thin continuum emission, $T_e^*$=10$^4$$\,$K and $N$(He$^+$)/$N$(H$^+$)=0.08.}
\tablenotetext{c}{ILTRs of the H41$\alpha$ and H31$\alpha$ RRLs measured toward MWC349A by \citet{mar89}.}
\tablenotetext{d}{ILTRs of the H34$\alpha$ RRL obtained toward MWC349A by \citet{thu92}.}

\end{deluxetable}

\section{Modelling of the extremely broad radio recombination line emission}
\label{mod}

\begin{figure}
\includegraphics[angle=0,scale=.55]{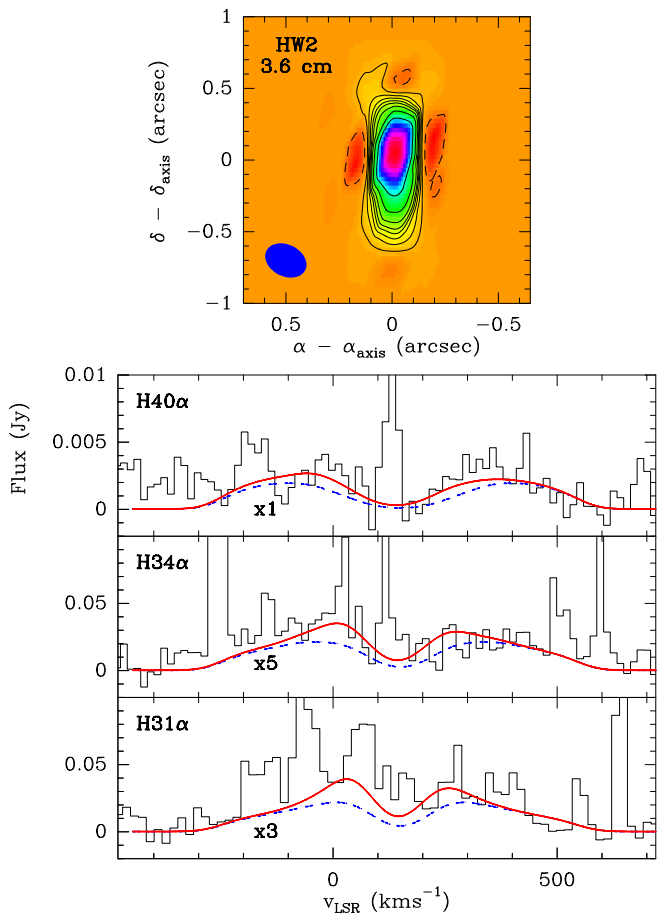}
\caption{{\it Upper panel:} Morphology of the
3.6$\,$cm radiocontinuum emission predicted by our model for 
the HW2 radio jet, smoothed to the angular 
resolution (0.25$"$$\times$0.18$"$) of the VLA observations 
of \citet{rod94}. The X and Y axis show the offsets 
(in arcseconds) with respect to the major and minor axis
of the jet. Contours are the same as in \citet{rod94}, 
i.e. -3, 3, 6, 9, 12, 15, 20, 30 and 50 times the rms 
(50$\,$$\mu$Jy) in the VLA 3.6$\,$cm image.
Beam is shown at the lower left corner.
{\it Lower panels:} Spectra of the H40$\alpha$, H34$\alpha$ 
and H31$\alpha$ lines predicted by our model assuming LTE 
(dashed lines) and non-LTE conditions (solid lines), 
overlaid on the RRL profiles measured toward HW2 (black histograms). 
Numbers indicate the factors by which the intensities of the predicted 
spectra have been multiplied in order 
to qualitatively compare the observations with the modeling (see Section$\,$\ref{mod}).\label{f2}}
\end{figure}

We try to reproduce the observed RRL profiles toward HW2 by using 
the 3D radiative transfer model of \citet[][2002, 2011]{mar93}, 
which includes the LTE departure coefficients, $b_n$ and 
$\beta_n$, calculated by \citet{wal90}. The model considers an 
isothermal ($\sim$ 10$^4$$\,$K) collimated, bi-conical ionized 
radio jet with a semi-opening angle of 18$^\circ$ and an 
inclination angle with respect to the line-of-sight of 52$^\circ$. 
This inclination angle is similar to that derived by 
\citet[][of 62$^\circ$$\pm$10$^\circ$]{pat05} or 
\citet[][of 56$^\circ$]{vle10}. 
However, the semi-opening angle (18$^\circ$) is a 
factor of $\geq$2 larger than that derived by 
\citet[][of $\sim$7.5$^\circ$]{rod94}. The 
determination of this angle is subject to large uncertainties 
since the radiocontinuum emission of the HW2 jet is variable, and 
its semi-minor axis is unresolved \citep{cur06}. Despite the 
18$^\circ$ semi-opening angle, our model reproduces well the radiocontinuum fluxes at centimeter wavelengths, and the extremely 
broad RRL profiles measured toward 
HW2 (see below). 

The radiocontinuum spectrum of HW2 is fitted with a mass loss rate of
3.2$\times$10$^{-6}$$\,$M$_\odot$$\,$yr$^{-1}$. Although
this rate is a factor of 4 larger than that estimated by 
\citet[][$\sim$8$\times$10$^{-7}$$\,$M$_\odot$$\,$yr$^{-1}$]{rod94}, 
the predicted radiocontinuum fluxes at centimeter wavelengths 
agree well (within 13\%) with those 
reported by these authors (Table$\,$\ref{tab3}). In 
Figure$\,$\ref{f2} (upper panel), we show the morphology of the
3.6$\,$cm (8.44$\,$GHz) radiocontinuum emission predicted by our model
for the HW2 radio jet. This image is very similar to that 
observed with the VLA \citep{rod94}. The 
derived free-free continuum fluxes at 3, 1.9 and 1.4$\,$mm 
(0.058, 0.079 and 0.094$\,$Jy, respectively) are also 
consistent with those shown in Table$\,$\ref{tab1}. 

\begin{deluxetable}{lccc}
\tablecaption{Comparison between the observed and predicted radiocontinuum 
fluxes toward Cepheus A HW2.\label{tab3}}
\tablewidth{0pt}
\tablehead{
\colhead{Frequency} & \colhead{Observed $S_{\nu}$\tablenotemark{a}} & \colhead{Predicted $S_{\nu}$\tablenotemark{b}} & \colhead{Relative error} \\
\colhead{(GHz)} & \colhead{(mJy)} & \colhead{(mJy)} & \colhead{(\%)}
}
\startdata

1.49 & 3.4 & 3.3 & 1.6\% \\ 
4.86 & 7.5 & 7.6 & 1.1\% \\ 
8.44 & 9.8 & 11.1 & 12.9\% \\
14.9 & 15.8 & 16.3 & 3.2\% \\
43.3 & 35 & 33.4 & 4.5\%

\enddata

\tablenotetext{a}{From \citet[][]{rod94}.}
\tablenotetext{b}{Predicted by our model for a mass loss rate
 of 3.2$\times$10$^{-6}$$\,$M$_\odot$$\,$yr$^{-1}$, and a semi-opening angle
of 18$^\circ$.}

\end{deluxetable}


The assumed electron density distribution in the HW2 jet 
varies as $r^{-2.11}$, with $r$ the distance to the central protostar
extending out to 0.41$"$ \citep[285$\,$AU at a distance of 700$\,$pc;][]{rei09}.
The electron density at the inner radius of 6.7$\,$AU is 
2.7$\times$10$^8$$\,$cm$^{-3}$, sufficient to obtain 
the maser effect in the RRLs. 

To explain the extremely broad linewidths and the lack of emission 
at systemic velocities, 
we have considered that the ionized gas in the jet is accelerated 
constantly to reach a terminal velocity of 500$\,$km$\,$s$^{-1}$ \citep{cur06} 
at 35$\,$AU from the protostar. The model also includes electron impact 
(pressure) broadening and a turbulent velocity of 60$\,$km$\,$s$^{-1}$.
However, pressure broadening is expected to be 
negligible for the observed RRLs \citep{ket08}.

Figure$\,$\ref{f2} (lower panels) shows the RRL profiles predicted
under LTE (dashed lines) and non-LTE conditions 
(solid lines), overlaid on the observed RRLs. To match the observations, we needed to (red-)shift the predicted RRLs by 160$\,$km$\,$s$^{-1}$ with 
respect to the systemic velocity of the cloud 
\citep[$-$10$\,$km$\,$s$^{-1}$;][]{mar05}. Water masers toward 
HW2 also show significant shifts 
in their peak velocities, associated with the dynamics of the 
HW2 rotating disk and wide-angle 
outflow \citep{torr96,torr11}. However, their velocity spans 
($\sim$40$\,$km$\,$s$^{-1}$) are too small to explain the 
160$\,$km$\,$s$^{-1}$ shift. The morphology and kinematics of 
the inner regions in the jet are unknown. Asymmetries in the 
density, temperature and kinematics between the red and blue 
lobes of the jet could give rise, due to the maser phenomenon, 
to large asymmetries in the RRL profiles, as expected
from the larger spatial scales needed for RRL maser 
amplification compared to molecular masers (see below). 
Future modelling will explore whether these asymmetries are responsible
for the 160$\,$km$\,$s$^{-1}$ shift observed in the RRLs.


From Figure$\,$\ref{f2}, we find that the model qualitatively 
reproduces the two-component line profiles, with zero-intensity 
linewidths of $\sim$1000$\,$km$\,$s$^{-1}$, of the RRLs toward HW2. 
It can be seen that maser amplification (solid lines) occurs at radial
velocities smaller than $\pm$300$\,$km$\,$s$^{-1}$ with respect 
to the systemic velocity of the jet. This is due to
i) high-enough electron densities at those velocities  
(3$\times$10$^5$-4$\times$10$^7$$\,$cm$^{-3}$; Strelnitski, Ponomarev \& Smith 1996); and ii) large enough {\it coherent lengths}, which lead to 
substantial amplification of the radiation along a velocity-coherent
path \citep{pon94}. 
For larger velocities ($\geq$$\pm$300$\,$km$\,$s$^{-1}$),
the electron densities are also high 
(3$\times$10$^5$-8$\times$10$^6$$\,$cm$^{-3}$). However, 
the coherence length is very small making the maser effect 
negligible \citep{pon94}. 
Maser amplification is maximum at radial velocities
of $\pm$140-200$\,$km$\,$s$^{-1}$ with respect to the systemic
velocity of the jet, because the electron densities 
at these velocities correspond to the optimum values for
maser amplification \citep[i.e. 
6.8$\times$10$^6$$\,$cm$^{-3}$ for H40$\alpha$, 
1.7$\times$10$^7$$\,$cm$^{-3}$ for H34$\alpha$, and 
3.0$\times$10$^7$$\,$cm$^{-3}$ for H31$\alpha$; 
Figure$\,$8 in][]{str96}. Toward MWC349A, \citet{mar94} also
reported the detection of RRL maser emission at velocities
very different from the ambient cloud velocity 
($\pm$60$\,$km$\,$s$^{-1}$), and predicted that they arise from 
ionized outflowing gas \citep{mar11}.

Although the model reproduces
well the intensity of the H40$\alpha$ line, it fails to
predict the intensities of the H34$\alpha$ and 
H31$\alpha$ lines (note that these intensities have been multiplied in 
Figure$\,$\ref{f2} by factors of 5 and 3, respectively). 
This problem of unmatched intensities for RRLs with n$<$40, was already 
noted by \citet{mar89} for the MWC349A RRL maser object, and could be due either to small density and temperature inhomogeneities in the stellar wind, or to uncertainties in the $\beta_n$ coefficients \citep{mar93}. Despite the unmatched intensities, our model 
reproduces the asymmetry observed in the RRLs 
(Section$\,$\ref{res}), which can only be explained by RRL maser 
effects. All this suggests that the extremely broad features
detected in the millimeter spectra toward Cepheus A HW2, are associated with RRL masers generated in the HW2 ionized jet. \\

In summary, we report the detection of the H40$\alpha$, H34$\alpha$ and
H31$\alpha$ RRLs toward the high-velocity ionized jet in Cepheus A HW2. This emission shows extremely broad line profiles with zero-intensity linewidths of $\sim$1100$\,$km$\,$s$^{-1}$. The derived ILTRs significantly deviate from those in LTE, suggesting that these lines 
are RRL masers formed in the HW2 ionized jet with a expanding 
velocity of 500$\,$km$\,$s$^{-1}$. 
Together with MWC349A, the discovery of RRL masers in HW2 suggests that 
this mechanism could be a common feature in dense UC HII regions \citep{mar02}.

\acknowledgments

IJ-S acknowledges the Smithsonian Astrophysical Observatory for the
support provided through a SMA fellowship. We would like to thank 
an anonymous referee who helped to improve the manuscript. 
JM-P and IJ-S have been 
partially funded by MICINN grants ESP2007-65812-C02-C01 and 
AYA2010-21697-C05-01 and AstroMadrid (CAM S2009/ESP-1496).


\end{document}